# Sedimentation-consolidation of a double porosity material

Henry Wong[a], Chin J. Leo[b], J.M. Pereira[c], Ph. Dubujet[d]

[a] Département Génie Civil et Bâtiment (URA CNRS 1652), Ecole Nationale des Travaux Publics de l'État, 2 rue Maurice Audin, 69518 Vaulx en Velin, France
[b] School of Engineering, University of Western Sydney, Locked Bag 1797 Penrith South DC, Sydney, NSW 1797, Australia
[c] Ecole Nationale des Ponts et Chaussées (ENPC), Institut Navier - CERMES, 6-8 av Blaise Pascal, 77455 Marne-la-Vallée cedex 2, France
[d] Ecole Nationale d'Ingénieur de St-Etienne (ENISE), 58, rue Jean PAROT - 42023 St-Etienne cedex 2, France

**Abstract.** This paper studies the sedimentation-consolidation of a double porosity material, such as lumpy clay. Large displacements and finite strains are accounted for in a multidimensional setting. Fundamental equations are derived using a phenomenological approach and non-equilibrium thermodynamics, as set out by Coussy [5]. These equations particularise to three non-linear partial differential equations in one dimensional context. Numerical implementation in a finite element code is currently being undertaken.

## 1. Introduction

It is quite common to find in porous materials an interweaving system of preferential pathways which facilitates the flow of fluid through the material. Fractured rock mass is an evident example, and lumpy clays used in land reclamation is another. In the earlier, the fractures are much more permeable to fluid flow than the intact host rock, while in the latter the inter-lump permeability of the clay is far higher than its intra-lump permeability. In these materials, the classic single porosity model for simulating flow has been found to be deficient, hence the concept of double porosity has been suggested by many investigators.

A double porosity model to describe flow in the fractured porous media in the petroleum industry was first proposed by Barenblatt *et al*. [2]. The fractured porous media was deemed to consist of two overlapping systems, one representing the fracture network and the other the porous blocks hence giving rise to the term "double porosity". As the porous media was assumed to be rigid, the coupling between fluid flow and mechanical effects was ignored. This coupling was later introduced by Aifantis [1]. The double porosity model was later adapted to the consolidation of *lumpy clay* in land reclamation works by Nogami *et al*. [10], who neglected the effects of self-weight, which is in principle the driving force of consolidation. Yang *et al*. [15] extended the theory to include the important effects of self-weight. These authors, however, assumed small displacements and small strains, which would appear unrealistic on account of the large change in porosity typically involved. The sedimentation process prior to consolidation was also not considered.

The present paper proposes an extension of the models developed previously to describe the sedimentation-consolidation behaviour of blocky soft porous materials such as lumpy clays based on the concept of double porosity. As such, the extension in the present paper unifies the sedimentation and consolidation processes of a double porosity material, and also includes formulation in Eulerian and Lagrangian frameworks to properly take into account the finite strains and displacements which previous studies have not undertaken. The model equations are obtained through a general multidimensional approach based on the principle of non-equilibrium thermodynamics and can be easily specialised to the case of small strains. Quite importantly, the proposed model considers the critical coupling effect due to deformation compatibility between the macro and micro pores within the system (i.e. preferential pathways and the porous lumps or blocks). This crucial coupling has been suggested by Khalili and Valliappan [6], Tuncay and Corapcioglu [17], Loret and Rizzi [9], Khalili *et al.* [7], Callari and Federico [4], Pao and Lewis [12] among others. Khalili [8] also has shown that neglecting this cross coupling effect can lead to spurious numerical results particularly in the early time response of the system.

Summation convention on repeated indices is adopted unless otherwise stated. Positive stresses and strains are taken to mean tensile stresses and elongations, while fluid pressures are taken positively.
lump

## 2. Consolidation zone: main assumptions and fundamental equations

Figure 1 shows a typical configuration of a soft blocky porous system, such as lumpy clay, for which the present model is applicable. The system of clay lumps saturated with water represents a double porous network. At a space scale an order of magnitude higher than the size of individual clay lumps (scale of a REV in figure 1), the material is idealised as a triphasic material, represented as the superposition of three continua [5], namely the intra-lump fluid phase "1", the inter-lump fluid phase "2" and the solid skeleton "s". Eulerian porosities $n_1$, $n_2$ and $n_s=1-n_1-n_2$ are defined such that, within a given elementary volume $d\Omega_t$ in the actual configuration, the volume of intra-lump fluid, inter-lump fluid and solid skeleton are respectively given by $n_1 d\Omega_t$, $n_2 d\Omega_t$ and $n_s d\Omega_t$. The sum $n=n_1+n_2$ gives the total Eulerian porosity. Each of the two fluid phases is assumed to form a continuous flow network and interact with each other through a mass exchange term. Each of the three phases has an independent trajectory described by velocity fields $\mathbf{v}_1$, $\mathbf{v}_2$ and $\mathbf{v}_s$. Since we are mainly interested by the solid skeleton, the description will be focused on the movement of the solid particles; the initial position at time t=0 (Lagrange coordinates) of a skeleton particle is denoted by $\mathbf{X}$ and its actual position at time t by $\mathbf{x}$ (Eulerian coordinates), the corresponding displacement being $\mathbf{U}=\mathbf{x}-\mathbf{X}$, with:

$$\mathbf{v}_s = \frac{d^s \mathbf{U}}{dt} = \dot{\mathbf{U}} \tag{1}$$

where $d^s/dt$ or a dot above a variable represents the material derivative with respect to the solid phase. We will use uppercase letters to denote vectorial operators in Lagrangian coordinates and small letters in Eulerian coordinates. Using this notation, the material derivative is defined as:

$$\dot{G} = \frac{d^s G}{dt} = \frac{\partial G}{\partial t} + \mathbf{v}_s \cdot \mathbf{grad} G \tag{2}$$

where **grad** is the gradient operator with respect to Eulerian coordinates $\mathbf{x}$. The second order deformation gradient tensor is denoted by:

$$\mathbf{F} = \mathbf{1} + \mathbf{Grad}(\mathbf{U}) \tag{3}$$

where **1** is the second order identity tensor and **Grad** the gradient operator with respect to Lagrangian coordinates $\mathbf{X}$. The Jacobian J of the transformation of the skeleton is defined as:

$$J = \det \mathbf{F} \quad ; \quad \frac{d^s J}{dt} = J \operatorname{div} \mathbf{v}_s \tag{4}$$

In other words, a skeleton particle initially of volume $d\Omega_0$ will occupy a volume $d\Omega_t = Jd\Omega_0$ at time t. The Green-Lagrange strain tensor, denoted by $\mathbf{\Delta}$, is related to J by:

$$\mathbf{\Delta} = (\,^t\mathbf{F}\cdot\mathbf{F} - \mathbf{1})/2 \quad ; \quad \dot{J} = J(\mathbf{F}^{-1}\cdot\,^t\mathbf{F}^{-1}):\dot{\mathbf{\Delta}} \tag{5}$$

The movement of the fluid phases is described by two Eulerian fluid mass fluxes relative to the solid skeleton, $\mathbf{w}^\alpha$:

$$\mathbf{w}^\alpha = n_\alpha \rho_f (\mathbf{v}_\alpha - \mathbf{v}_s) \quad ; \quad \alpha = 1, 2 \text{ (no summation over } \alpha) \tag{6}$$

We will suppose in the sequel that the fluid filling up the intra-lump and inter-lump voids is incompressible so that the fluid mass density $\rho_f$ is constant. Based on the definition of the Jacobian J, we also define the Lagrangian porosities:

$$\phi_\alpha = Jn_\alpha \ (\alpha = 1, 2) \quad ; \quad \phi = \phi_1 + \phi_2 \quad ; \quad \phi = Jn \tag{7}$$

such that $\phi_\alpha d\Omega_0 = n_\alpha d\Omega_t$ is the volume of fluid phase $\alpha$ in $d\Omega_t$. The Lagrangian fluid mass fluxes $\mathbf{M}^\alpha$ are defined such that $\mathbf{M}^\alpha \cdot \mathbf{N}dA = \mathbf{w}^\alpha \cdot \mathbf{n}da$ (which implies $\text{Div}\mathbf{M}^\alpha = J\text{div}\mathbf{w}^\alpha$), where da with unit normal $\mathbf{n}$ is the oriented material surface at time t which measures dA with unit normal $\mathbf{N}$ at t=0. Recalling that $\mathbf{n}da = J\,^t\mathbf{F}^{-1}\cdot\mathbf{N}dA$, we have:

$$\mathbf{M}^\alpha = J\,\mathbf{F}^{-1}\cdot\mathbf{w}^\alpha \tag{8}$$

## 3. Mass balance equations

Taking into account the fluid incompressibility, the mass balance of fluid phases writes:

$$\frac{\partial n_\alpha}{\partial t} + \text{div}(n_\alpha \mathbf{v}_\alpha) = \hat{m}_{\to\alpha} J^{-1} \rho_f^{-1} \quad ; \quad \alpha = 1, 2 \text{ (no summation over } \alpha) \tag{9}$$

where $\hat{m}_{\to\alpha} d\Omega_0 dt$ is the fluid mass increase of phase $\alpha$ inside the overall current volume $d\Omega_t$ during time dt, due to exchanges between the two fluid phases. Overall fluid mass conservation implies:

$$\hat{m}_{\to 1} + \hat{m}_{\to 2} = 0 \tag{10}$$

The mass of solid skeleton contained inside an elementary volume $d\Omega_t$ following the movement of solid skeleton is invariant, hence:

$$m_s d\Omega_0 = \rho_s (J - \phi) d\Omega_0 = \rho_s (1 - n) d\Omega_t = \rho_s (1 - \phi_0) d\Omega_0 \tag{11}$$

Here $m_s$ and $\rho_s$ are respectively the solid mass content per unit initial volume and the density, taken to be constants for incompressible solid phase. We thus deduce that:

$$m_s / \rho_s = J - \phi = (1 - n)J = (1 - \phi_0) \quad ; \quad J = (1 - \phi_0)(1 - n)^{-1} = 1 + \phi - \phi_0 \tag{12}$$

The fluid mass of phase $\alpha$ inside an elementary volume $d\Omega_t$ may be expressed via another Lagrangian variable as $m_\alpha d\Omega_0 = \rho_f \phi_\alpha d\Omega_0 = \rho_f n_\alpha d\Omega_t$, hence:

$$m_\alpha = \rho_f \phi_\alpha = J\rho_f n_\alpha \tag{13}$$

is the mass content of fluid phase $\alpha$ per unit initial overall volume. The fluid mass conservation relation (9) can also be expressed in terms of Lagrangian variables, using (7), (8) and (13):

$$\dot{m}_\alpha + \text{Div}\mathbf{M}^\alpha = \hat{m}_{\to\alpha} \tag{14}$$

**4. Momentum balance equations**

Neglecting acceleration terms, the momentum balance in terms of Eulerian variables writes:

$$\text{div}\boldsymbol{\sigma} + \rho\mathbf{g} - J^{-1}\hat{m}_{\to 2}(\mathbf{v}_2 - \mathbf{v}_1) = \mathbf{0} \quad ; \quad \rho = (1-n)\rho_s + n\rho_f \tag{15}$$

where $\rho$ is overall density of the mixture and $\mathbf{g}$ is the gravity. In Lagrangian variables, we have:

$$\text{Div}(\mathbf{F}\cdot\boldsymbol{\pi}) + (m_s + m_f)\mathbf{g} - \hat{m}_{\to 2}(\mathbf{v}_2 - \mathbf{v}_1) = \mathbf{0} \quad ; \quad m_f = m_1 + m_2 \tag{16}$$

Recall that the Piola-Kirchoff stress tensor $\boldsymbol{\pi}$ and Cauchy stress tensor $\boldsymbol{\sigma}$ are related by :

$$\boldsymbol{\sigma} = \frac{1}{J}\mathbf{F}\cdot\boldsymbol{\pi}\cdot{}^t\mathbf{F} \tag{17}$$

**5. Thermodynamics analysis**

Neglecting the acceleration terms (see [5] and [13] for details), the dissipation per unit current overall volume can be decomposed into four components (skeleton, phase change, mass and heat transfer):

$$\varphi = \varphi_s + \varphi_{\to} + \varphi_t + \varphi_h$$

$$= \boldsymbol{\sigma}:\text{grad}(\mathbf{v}_s) + \mu^\alpha \dot{m}_\alpha J^{-1} - s\dot{T} - \dot{\psi} - \psi\,\text{div}(\mathbf{v}_s) + \frac{\hat{m}_{\to 2}}{J}\left[\left(\mu^1 + \frac{(\mathbf{v}_1 - \mathbf{v}_s)^2}{2}\right) - \left(\mu^2 + \frac{(\mathbf{v}_2 - \mathbf{v}_s)^2}{2}\right)\right] \tag{18}$$

$$+ \sum_\alpha \left[-s^\alpha \text{grad}T - \text{grad}\mu^\alpha + \mathbf{g}\right]\cdot\mathbf{w}^\alpha + \left(-\frac{1}{T}\mathbf{q}\cdot\text{grad}T\right) \geq 0$$

where $\mu^\alpha$ and $s^\alpha$ are the mass-based chemical potential and entropy of fluid phase "$\alpha$" and $\psi$ the total free energy per unit current volume. Multiplying by the Jacobian J, and denoting the Lagrangian densities $S = Js$, $\Psi = J\psi$ and $\Phi_\pi = J\varphi_\pi$ (ie. per unit initial volume), the Lagrangian equivalence of (18) is:

$$\Phi = \Phi_s + \Phi_{\to} + \Phi_t + \Phi_h$$

$$= \boldsymbol{\pi}:\dot{\boldsymbol{\Delta}} + \mu^\alpha \dot{m}_\alpha - S\dot{T} - \dot{\Psi} + \hat{m}_{\to 2}\left[\left\{\mu^1 + \frac{(\mathbf{v}_1 - \mathbf{v}_s)^2}{2}\right\} - \left\{\mu^2 + \frac{(\mathbf{v}_2 - \mathbf{v}_s)^2}{2}\right\}\right] \tag{19}$$

$$+ \sum_\alpha \left[-s^\alpha \text{Grad}T - \text{Grad}\mu^\alpha + \mathbf{g}\cdot\mathbf{F}\right]\cdot\mathbf{M}^\alpha + \left(-\frac{1}{T}\mathbf{Q}\cdot\text{Grad}\right) \geq 0$$

where we have accounted for $2\dot{\boldsymbol{\Delta}} = {}^t\mathbf{F}\cdot(\text{grad}\mathbf{v}_s + {}^t\text{grad}\mathbf{v}_s)\cdot\mathbf{F}$ and $\boldsymbol{\pi} = J\mathbf{F}^{-1}\cdot\boldsymbol{\sigma}\cdot{}^t\mathbf{F}^{-1}$. From now on, we will assume isothermal conditions $dT = 0$ (hence $\varphi_h = 0$). We will adopt herein the classic decoupling of the various components of dissipation and will require each of them to be independently non-negative. Recalling the following classic thermodynamics relations of fluids:

$$\psi_\alpha = e_\alpha - Ts_\alpha \quad ; \quad \mu^\alpha = \psi_\alpha + P_\alpha/\rho_f \quad ; \quad de_\alpha = -P_\alpha d(1/\rho_f) + Tds_\alpha \qquad (20)$$

Here, $P_\alpha$, $\psi_\alpha$ and $e_\alpha$ are respectively the fluid pressure, the free and internal energy per unit mass of fluid phase "$\alpha$". From (20), we deduce that:

$$d\psi_\alpha = -P_\alpha d\left(\frac{1}{\rho_f}\right) - s_\alpha dT \quad ; \quad d\mu_\alpha = \frac{dP_\alpha}{\rho_f} - s_\alpha dT \qquad (21)$$

Under isothermal conditions, $dT = 0$, we get:

$$d\mu^\alpha = dP_\alpha/\rho_f \qquad (22)$$

In view of the last result, the non-negativity of $\varphi_t$ can be satisfied by the generalised Darcy's law:

$$\mathbf{w}^\alpha = -k_\alpha [\mathbf{grad} P_\alpha - \rho_f \mathbf{g}] \qquad (23)$$

where $k_\alpha = K_\alpha/\rho_\alpha$, and $K_\alpha$ (in m/s) is the classic hydraulic conductivity. The non-negativity of $\varphi_\rightarrow$ (ou $\Phi_\rightarrow$), neglecting the relative velocities and on account of (22), leads to the inequality:

$$\hat{m}_{\rightarrow 2}(P_1 - P_2) \geq 0 \qquad (24)$$

which can be satisfied by the following phenomenological relation

$$\hat{m}_{\rightarrow 2} = -\hat{m}_{\rightarrow 1} = \omega(P_1 - P_2) \qquad (25)$$

where the positive constant $\omega$ is the exchange coefficient. Note that the above relation has already been suggested by [5], [6] and [15] in an empirical manner. Our theoretical development shows that this relation can be derived naturally from thermodynamic principles. We now decompose the total free energy into the sum of the skeleton free energy and the fluid free energies $\Psi = \Psi_s + m_\alpha \psi_\alpha$. Substituting this into the expression of the dissipation above, and on account of (20), we get:

$$\Phi_s = \boldsymbol{\pi} : \dot{\boldsymbol{\Delta}} + P_\alpha \dot{\phi}_\alpha - \dot{\Psi}_s \geq 0 \qquad (26)$$

This form suggests that the natural independent variables of $\Psi_s$ are $\boldsymbol{\Delta}$, $\phi_\alpha$ and $\chi$, so that:

$$\Phi_s = (\boldsymbol{\pi} - \frac{\partial \Psi_s}{\partial \boldsymbol{\Delta}}) : \dot{\boldsymbol{\Delta}} + \sum_\alpha (P_\alpha - \frac{\partial \Psi_s}{\partial \phi_\alpha})\dot{\phi}_\alpha - \frac{\partial \Psi_s}{\partial \chi}\dot{\chi} \geq 0 \qquad (27)$$

where $\chi$ is the internal variable used to describe irreversabilities. This leads to the classic state equations:

$$\boldsymbol{\pi} = \frac{\partial \Psi_s}{\partial \boldsymbol{\Delta}} \quad ; \quad P_\alpha = \frac{\partial \Psi_s}{\partial \phi_\alpha} \qquad (28)$$

and the remaining dissipation term:

$$\Phi_s = -\frac{\partial \Psi_s}{\partial \chi}\dot{\chi} \geq 0 \qquad (29)$$

Assuming reversible behaviour so that $\Phi_s = 0$ and the dependence on the internal variables $\chi$ disappears, the second order cross derivatives being independent of the order of differentiation leads to the Maxwell equations:

$$\frac{\partial P_\alpha}{\partial \Delta_{ij}} = \frac{\partial^2 \Psi_s}{\partial \Delta_{ij} \partial \phi_\alpha} = \frac{\partial^2 \Psi_s}{\partial \phi_\alpha \partial \Delta_{ij}} = \frac{\partial \pi_{ij}}{\partial \phi_\alpha} = B_{ij}^\alpha \qquad (30)$$

Differentiating $\pi_{ij}$ and $P_\alpha$ then gives:

$$\dot{\pi}_{ij} = L'_{ijkl} \dot{\Delta}_{kl} + B_{ij}^\alpha \dot{\phi}_\alpha \qquad (31)$$

$$\dot{P}_\alpha = B_{ij}^\alpha \dot{\Delta}_{ij} + r_{\alpha k} \dot{\phi}_k \qquad (32)$$

with $\quad L'_{ijkl} = \frac{\partial^2 \Psi_s}{\partial \Delta_{ij} \partial \Delta_{kl}} \; ; \; r_{\alpha k} = \frac{\partial^2 \Psi_s}{\partial \phi_\alpha \partial \phi_k}$.

Using the Legendre transformation $G_s = \Psi_s - P_\alpha \phi_\alpha$ and substituting into (26), we can show that:

$$\dot{\pi}_{ij} = L_{ijkl} \dot{\Delta}_{kl} - b_{ij}^\alpha \dot{P}_\alpha \qquad (33)$$

$$\dot{\phi}_\alpha = b_{ij}^\alpha \dot{\Delta}_{ij} + R_{\alpha k} \dot{P}_k \qquad (34)$$

$$L_{ijkl} = \frac{\partial^2 G_s}{\partial \Delta_{ij} \partial \Delta_{kl}} \quad ; \quad -\frac{\partial \phi_\alpha}{\partial \Delta_{ij}} = \frac{\partial^2 G_s}{\partial \Delta_{ij} \partial P_\alpha} = \frac{\partial \pi_{ij}}{\partial P_\alpha} = -b_{ij}^\alpha \quad ; \quad R_{\alpha k} = -\frac{\partial^2 G_s}{\partial P_\alpha \partial P_k} \qquad (35)$$

For classic isotropic behaviour, we have $L_{ijkl} = \mu(\delta_{ik}\delta_{jl} + \delta_{il}\delta_{jk}) + \lambda \delta_{ij}\delta_{kl}$ and $b_{ij}^\alpha = b_\alpha \delta_{ij}$, hence:

$$\dot{\pi}_{ij} = (2\mu \delta_{ik}\delta_{jl} + \lambda \delta_{ij}\delta_{kl})\dot{\Delta}_{kl} - b_\alpha \dot{P}_\alpha \delta_{ij} \qquad (36)$$

$$\dot{\phi}_\alpha = b_\alpha \dot{\Delta}_{mm} + R_{\alpha k} \dot{P}_k \qquad (37)$$

where $b_\alpha$ is *a priori* positive as a positive pressure increment should lead to an algebraic stress reduction in (36), due to the traction-positive sign convention. Through the Maxwell symmetry relation (35), this leads to the consistent result that an overall volume increase will lead to increase in porosities according to (37). Note that constitutive relations (36) and (37) are also consistent with the previous work of Khalili & Villiapan [6] under small strains. Their analytical formulae therefore furnish a first estimate for the parameters $b_\alpha$ and $R_{\alpha k}$ under large strains. Equation (36) can be rewritten as:

$$(\pi_{ij} + b_\alpha P_\alpha \delta_{ij})^{\cdot} = L_{ijkl}\dot{\Delta}_{kl} = (2\mu \delta_{ik}\delta_{jl} + \lambda \delta_{ij}\delta_{kl})\dot{\Delta}_{kl} \qquad (38)$$

suggesting the following effective stress rate:

$$\pi'_{ij} = \pi_{ij} + b_\alpha P_\alpha \delta_{ij} \qquad (39)$$

## 7. One dimensional consolidation under oedometric conditions

In one dimensional oedometric conditions, only the displacement in the direction $x_1 = x$ ($X_1 = X$) is non zero. Similarly only the strain component $\Delta_{11} = \Delta$ is non zero. Denoting $\sigma_{11} = \sigma$ and $\pi_{11} = \pi$, we deduce from (17) and (8) that $\sigma = J\pi$ and $w^\alpha = M^\alpha$. Neglecting the momentum contribution $\hat{m}_{\to\alpha}(\mathbf{v}_2 - \mathbf{v}_1)$, equations (14), (15), (16) and (23) simplifies to:

$$\dot{m}_\alpha + \frac{\partial M^\alpha}{\partial X} = \hat{m}_{\to\alpha} \tag{40}$$

$$\frac{\partial \sigma}{\partial x} - \rho g = 0 \tag{41}$$

$$\frac{\partial (J\pi)}{\partial X} - (m_s + m_f)g = 0 \tag{42}$$

$$w^\alpha = M^\alpha = -k_\alpha\left(\frac{\partial P_\alpha}{\partial x} + \rho_f g\right) = -k_\alpha\left(\frac{\partial P_\alpha}{\partial X}J^{-1} + \rho_f g\right) \quad ; \quad \alpha=1, 2 \text{ (no summation over } \alpha) \tag{43}$$

$$\dot{\phi}_\alpha = b_\alpha \dot{\Delta} + R_{\alpha k}\dot{P}_k \tag{44}$$

Following [4], the constitutive law for the skeleton can be written as:

$$\pi' = \pi + b_\alpha P_\alpha = E\varpi(\phi) \tag{45}$$

where E is a reference oedometric modulus. We will assume in the sequel that the clay lumps come into contact with each other and consolidation starts when the porosity $\phi$ falls below a critical value $\phi_c$. Hence, $\varpi(\phi)$ will be zero for $\phi > \phi_c$, and starts to climb up when $\phi$ falls below $\phi_c$. Equations (1), (4), (5) and (12) allow to deduce the following relations in one dimensional cases:

$$\Delta = \frac{1}{2}(J^2 - 1) \; ; \; \dot{\Delta} = J\dot{J} \; ; \; J = 1 + \frac{\partial U}{\partial X} \; ; \; \dot{J} = \frac{\partial^2 U}{\partial X \partial t} \; ; \; \frac{\partial J}{\partial X} = \frac{\partial^2 U}{\partial X^2} \; ; \; \phi = \frac{\partial U}{\partial X} + \phi_0 \; ; \; \frac{\partial \phi}{\partial X} = \frac{\partial^2 U}{\partial X^2} + \frac{\partial \phi_0}{\partial X} \tag{46}$$

These relations together with (37) allow J, $\phi_\alpha$, $n_\alpha$ to be expressed in terms of U. Substitution of (13), (25), (43) and (46) into (40), we get 2 equations:

$$\rho_f\left(b_1\left(1 + \frac{\partial U}{\partial X}\right)\frac{\partial^2 U}{\partial X \partial t} + R_{11}\dot{P}_1 + R_{12}\dot{P}_2\right) - \frac{\partial}{\partial X}\left[k_1\left(\frac{\partial P_1}{\partial X}\left(1 + \frac{\partial U}{\partial X}\right)^{-1} + \rho_f g\right)\right] = -\omega(P_1 - P_2) \tag{47}$$

$$\rho_f\left(b_2\left(1 + \frac{\partial U}{\partial X}\right)\frac{\partial^2 U}{\partial X \partial t} + R_{21}\dot{P}_1 + R_{22}\dot{P}_2\right) - \frac{\partial}{\partial X}\left[k_2\left(\frac{\partial P_2}{\partial X}\left(1 + \frac{\partial U}{\partial X}\right)^{-1} + \rho_f g\right)\right] = \omega(P_1 - P_2) \tag{48}$$

A third equation is obtained by developing the momentum balance equation (42), using (45):

$$\left(1 + \frac{\partial U}{\partial X}\right)\left[E\frac{\partial \varpi(\phi)}{\partial X} - \frac{\partial (b_\alpha P_\alpha)}{\partial X}\right] + (E\varpi(\phi) - b_\alpha P_\alpha)\frac{\partial^2 U}{\partial X^2} - \left[m_s + \rho_f(\frac{\partial U}{\partial X} + \phi_0)\right]g = 0 \tag{49}$$

The three partial differential equations (47), (48) and (49), together with the adequate boundary conditions, constitute a three-fields-problem on (U, $P_1$, $P_2$). Dependency of $k_\alpha$ on the Lagrangian porosity $\phi_\alpha$ (say $k_\alpha = k_{\alpha 0}\delta(\phi_\alpha)$) can easily be accounted for, using:

$$\frac{\partial k_\alpha(\phi_\alpha)}{\partial X} = k_{\alpha 0} \frac{d\delta(\phi_\alpha)}{d\phi_\alpha} \frac{\partial \phi_\alpha}{\partial X} = k_{\alpha 0} \frac{d\delta(\phi_\alpha)}{d\phi_\alpha} \beta_\alpha \left( \frac{\partial^2 U}{\partial X^2} + \frac{\partial \phi_0}{\partial X} \right) \qquad (50)$$

It is more usual to make $k_\alpha$ depend on the Eulerian porosity $n_\alpha$ (say $k_\alpha = k_{\alpha 0}\delta(n_\alpha)$), the calculations will be slightly more complicated but no theoretical difficulties arise.

$$\frac{\partial k_\alpha(n_\alpha)}{\partial X} = k_{\alpha 0}(\phi_\alpha) \frac{d\delta(n_\alpha)}{dn_\alpha} \frac{\partial n_\alpha}{\partial X} = k_{\alpha 0} \frac{d\delta(n_\alpha)}{dn_\alpha} \beta_\alpha J^{-2} \left( (1-\phi_0) \frac{\partial^2 U}{\partial X_2} + J \frac{\partial \phi_0}{\partial X} \right) \qquad (51)$$

One example of the relative permeability function $\delta(n_\alpha)$ is the Kozeny-Carman relation [4]:

$$\delta(n_\alpha) = \left( \frac{n_\alpha}{n_{\alpha 0}} \right)^3 \left( \frac{1-n_{\alpha 0}}{1-n_\alpha} \right)^2 \qquad (52)$$

applicable to granular materials where $n_{\alpha 0}$ is the value of the Eulerian porosity at a given reference state (at which $k_\alpha = k_{\alpha 0}$). As for soft clayey materials, the following function between void ratio and logarithmic relative permeability often applies:

$$e_\alpha - e_{\alpha 0} = c_k \log_{10}(\delta) \qquad (53)$$

where $c_k$ is slope of the $e_\alpha$-$\log_{10}(k_\alpha)$ relationship (e.g. Tavenas et al. [16]), $e_{\alpha 0}$ is the void ratio corresponding to $n_{\alpha 0}$, and the void ratio and the porosity are related by the following expression: $n_\alpha = e_\alpha/(1+e_\alpha)$.

## 8. Sedimentation zone

In the sedimentation zone, the clay lumps are not in contact and therefore the intra-lump voids do not form a continuous flow network. The double porosity model described above cannot simulate this regime consistently. In keeping with the sedimentation analysis of [5], we will suppose:

(H1) No consolidation of clay lumps takes place in the sedimentation zone and the intra-lump voids remain constant, in other words $\phi_1 = \phi_{10}$. Hence $M^1 = w^1 = 0$.

(H2) $v_1 = v_s$, the clay lumps move as undeformable solid grains through the inter-lump fluid. They appear at the macroscopic scale as grains falling like rigid bodies with apparent density:

$$\rho'_s = \frac{(1-\phi_{10}-\phi_{20})\rho_s + \rho_f \phi_{10}}{1-\phi_{20}} \qquad (54)$$

(H3) The effective stress within clay lumps is zero, which implies $P_1 = P_2$, noted P. There is no contact between clay lumps hence the overall effective stress is also zero, thus $\sigma' = \sigma + P = 0$, implying:

$$\sigma = -P \qquad (55)$$

The mass conservation of each phase {s, 1, 2} with no mass exchange implies:

$$\frac{\partial(1-n_1-n_2)}{\partial t} + \frac{\partial}{\partial x}((1-n_1-n_2)v_s) = 0 \quad ; \quad \frac{\partial n_1}{\partial t} + \frac{\partial}{\partial x}(n_1 v_1) = 0 \quad ; \quad \frac{\partial n_2}{\partial t} + \frac{\partial}{\partial x}(n_2 v_2) = 0 \qquad (56)$$

Their sum gives: $\partial v_G / \partial x = 0$ with $v_G = (1-n)v_s + n_1 v_1 + n_2 v_2$ implying that $v_G$ is constant inside the sedimentation zone. The same conclusion can be reached by adding the 3 mass conservation equations in the consolidation zone. In this latter zone, the mass exchange between fluid phases 1 and 2 is non-zero but they cancel out each other, leading to the same conclusion that $v_G$ is constant in the consolidation zone. Since at the bottom (X=x=0), all velocities vanish, we therefore conclude that $v_G = 0$ everywhere (see for example [3]). Moreover, equations (40) to (42) hold with $\hat{m}_{\to\alpha} = 0$, while (43) only holds for α=2. Combining (41) and (55) gives:

$$\frac{\partial P}{\partial x} = -\rho g \tag{57}$$

Substitution into (43) for α=2 yields:

$$w^2 = M^2 = k_2(\rho - \rho_f)g = k_2(1-n)(\rho_s - \rho_f)g \tag{58}$$

The time derivative of (13), on account of (H1) and (12) leads to:

$$\dot{m}_2 = \rho_f \dot{\phi}_2 = \rho_f \dot{\phi} = \rho_f (1-\phi_0)\frac{d^s}{dt}\left(\frac{1}{1-n}\right) \tag{59}$$

Substitution of (58) and (59) into (40), with $\hat{m}_{\to 2} = 0$ finally leads to the equation sought:

$$\rho_f(1-\phi_0)\frac{\partial}{\partial t}\left(\frac{1}{1-n}\right)_{X=cte} + (\rho_s - \rho_f)g\frac{\partial}{\partial X}\{k_2(1-n)\} = 0 \tag{60}$$

By inspection, we find that if $\phi_2$ is independent of space and time, then $\dot{\phi} = \dot{J} = 0$. This implies that J=1, and $n_2=\phi_2$ are constant, so is $k_2$, hence the above equation is identically satisfied. Therefore, $\phi_2$=cte is a particular solution in the sedimentation zone. This would indeed be the case if initially the clay lumps are dispersed in a homogeneous manner, with a uniform porosity $\phi_{20}$. We will restrict ourselves hereafter to this particular case in order to compare the present formulation with previous works. With $v_G=0$ and (H2), we deduce that $v_s = -n_2(v_2-v_s)$. The right hand member can be explicitly determined using (6) and (58). This allows to deduce the solid skeleton velocity and subsequently the displacement in the sedimentation zone:

$$v_s^{sed} = -k_{20}\frac{(1-n_0)}{\rho_f}(\rho_s - \rho_f)g \quad ; \quad U = v_s^{sed} t \tag{61}$$

In other words, the upper boundary of the sedimentation zone goes down with a constant speed of $v_s^{sed}$, so that:

$$x_t(t) = x(H,t) = H + v_s^{sed} t \tag{62}$$

Above $x_t(t)$ is the clear water zone in which no solid particle is present, hence the fluid pressure inside the sedimentation zone $x_c(t) < x < x_t(t)$ is given by:

$$P_1 = P_2 = P = \rho_f g(H - x_t(t)) + \rho_0 g(x_t(t) - x) \tag{63}$$

with $\rho_0 = (1-\phi_0)\rho_s + \phi_0 \rho_f$ the initial overall density of the double porous media.

## 9. Sedimentation-consolidation interface

The sedimentation and consolidation zones are separated by a moving interface $x_c(t)$ of which the Lagrangian counterpart is $X_c(t)$. The interface advancement speed is governed by a jump condition which results from fluid mass conservation [5]:

$$[\![M - m_f C]\!]_{X_c} = 0 \tag{64}$$

where C is the Lagrangian speed of advancement of the discontinuity and $[\![f]\!]_X = f(X^+) - f(X^-)$ denotes the jump of a function across the position X. Classical results like [3], [5], [11] and [14] show that $\phi_2$ has a jump-discontinuity. In our case, $\phi_1$ should be continuous as the effective stress "just" starts to become non-zero at $X_c^-$ so as to induce consolidation of the clay lumps. On the sedimentation side $X_c^+$:

$$M = M^2 = k_{20}(1-n_0)(\rho_s - \rho_f)g \quad ; \quad m_f = \rho_f \phi = \rho_f(\phi_{10} + \phi_{20}) \tag{65}$$

while on the consolidation side $X_c^-$:

$$M = M^1 + M^2 \; ; \; M^\alpha = -k_\alpha(\phi_\alpha)\left(\frac{\partial P_\alpha}{\partial X}J^{-1} + \rho_f g\right) \; ; \quad m_f = \rho_f \phi = \rho_f(\phi_{10} + \phi_{2c}) \tag{66}$$

Substitution of (65) and (66) into (64) yields the Lagrangian advancement speed of the interface:

$$C = \frac{dX_c}{dt} = \frac{k_1(\phi_{10})\left(\frac{\partial P_1}{\partial X}J^{-1} + \rho_f g\right) + k_2(\phi_{2c})\left(\frac{\partial P_2}{\partial X}J^{-1} + \rho_f g\right) + k_{20}(1-n_0)(\rho_s - \rho_f)g}{\rho_f(\phi_{20} - \phi_{2c})} \tag{67}$$

## 10. Continuity and boundary conditions

Equations (47) - (49) need to be solved with adequate boundary conditions. At the interface $x_c(t)$, the three unknowns (U, $P_1$, $P_2$) are continuous, given by (61) and (63). At the bottom X=x=0, the impervious base condition writes:

$$\frac{\partial P_\alpha}{\partial X}J^{-1} + \rho_f g = 0 \quad (\alpha = 1, 2) \tag{68}$$

while the displacement U must satisfy:

$$U = 0 \tag{69}$$

Owing to the highly non-linear character of the system of equations involved, they have to be solved numerically.

## 11. Conclusions

A theoretical model describing the sedimentation-consolidation of a soft double porosity material such as lumpy clay has been introduced starting from a non-equilibrium thermodynamics approach. The formulation includes the coupling effect due to deformation compatibility between the macro and micro pores within the double porosity system. Both Eulerian and Lagrangian descriptions are used in this model, taking into account large displacements and finite strains. Equations of general validity are obtained in a multi-dimensional setting, before particularising to the case of one dimension. In this case, the theoretical developments result in a system of three partial differential equations to describe the large strain consolidation and one partial differential equation to describe the sedimentation, with an additional equation on the interface position. All equations are highly non-linear and therefore can only be solved numerically. Their numerical implementation into a finite element code is on-going. We leave the details of the variational formulation, finite element discretisation and numerical results to a later paper.

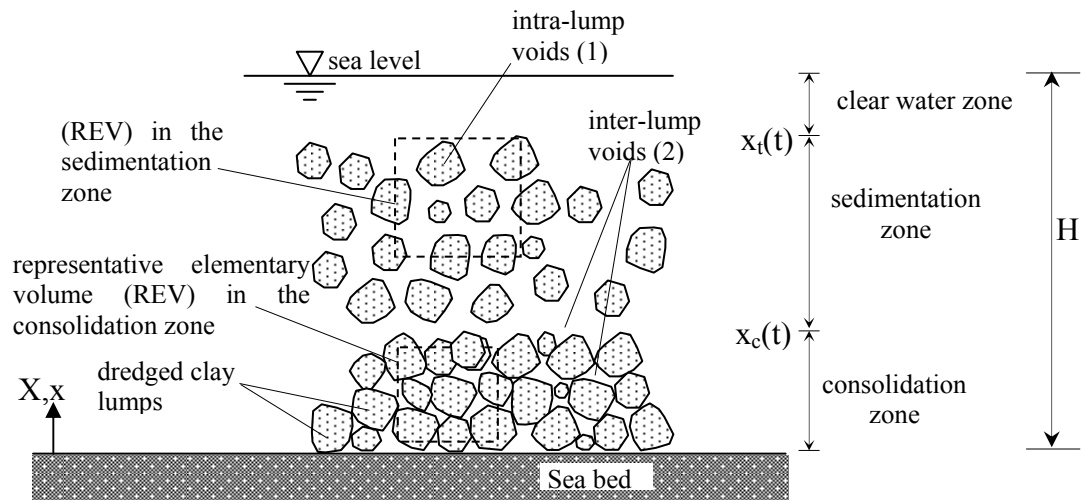

**Figure 1**. Conceptual representation of the sedimentation-consolidation in a lumpy clay.